\documentclass[prb,aps,preprint,showpacs,preprintnumbers,amsmath,amssymb,secnumroman]{revtex4}

\usepackage{amsmath}
\usepackage{amssymb}
\usepackage{graphicx}
\usepackage{dcolumn}
\usepackage{bm}

\newcommand{\beq}{\begin{equation}}
\newcommand{\eeq}{\end{equation}}
\newcommand{\beqa}{\begin{eqnarray}}
\newcommand{\eeqa}{\end{eqnarray}}
\newcommand{\vc}[1]{\mbox{\boldmath $#1$}}

\newcommand{\vol}[1]{{\bf #1}}

\begin{document}


\title{Use of the Basset coefficient in the calculation of the velocity of a spheroid slowing down in a viscous incompressible fluid after a sudden impulse}

\author{B. U. Felderhof}

 \email{ufelder@physik.rwth-aachen.de}
\affiliation{Institut f\"ur Theoretische Physik A\\ RWTH Aachen University\\
Templergraben 55\\52056 Aachen\\ Germany\\
}%

\date{\today}

\begin{abstract}
The motion of a spheroid in a viscous incompressible fluid after a sudden small impulse in the direction of the symmetry axis is studied on the basis of the linearized Navier-Stokes equations. The time-dependence of the spheroid velocity follows by a Fourier transform from the frequency-dependence of the impedance involving friction coefficient, body mass, and added fluid mass. A term proportional to the square root of frequency in the asymptotic high frequency expansion of the impedance, with a Basset coefficient, describes the initial decay in time from the initial value determined by the effective mass. It is shown from numerical evidence based on the exact multipole expansion of the solution for the flow field in terms of spheroidal wavefunctions that the knowledge of the Basset coefficient is insufficient for a reliable estimate of the deviations from a simple two-pole approximation to the complete behavior in time. The two-pole approximation can be calculated from the effective mass and the steady state friction coefficient.
\end{abstract}

\maketitle

\section{\label{I}Introduction}

The motion of a spheroid in a viscous incompressible fluid, after a sudden small impulse directed along the symmetry axis, may in principle be calculated from the linearized Navier-Stokes equations for fluid velocity and pressure, in combination with a boundary condition for the flow velocity at the surface of the spheroid. In the following we assume that the no-slip boundary condition holds. For a small body the no-slip condition must be relaxed and replaced by a partial slip boundary condition \cite{1}.

The added mass tensor, which together with the body mass determines the initial velocity after the impulse, has long been known from the general expressions for an ellipsoid \cite{2}. Similarly, the steady state friction tensor, which determines the steady state velocity for a constant force, is also known for the general ellipsoid \cite{2}. The complete time-dependence of the velocity after a sudden impulse follows in principle by Fourier transform from the frequency-dependence of the friction tensor, but this is not known for a general ellipsoid. For axial motion of a spheroid the frequency-dependent friction coefficient can be found in principle from a set of coupled equations for the multipole moments of the force density induced on the spheroid, as shown first by Lai and Mockros \cite{3}. It was shown by Lawrence and Weinbaum \cite{4}, that the equations can be solved efficiently by truncation and matrix inversion.

At low frequency the friction coefficient for axial motion is linear in the square root of frequency, with a coefficient which can be found from the friction coefficient at zero frequency \cite{5}$^,$\cite{6}. As a consequence of this relation the velocity relaxation function shows a $t^{-3/2}$ long-time tail, independent of shape, size, and mass density of the spheroid. The transient effects on the short and intermediate time scale are less well known.

For the case of a sphere, Stokes \cite{7} derived a simple explicit formula for the frequency-dependent friction coefficient. The impedance of a sphere is quadratic in the square root of frequency. The term linear in the square root is relevant both at low and at high frequency. The corresponding motion of a sphere has been studied by Boussinesq \cite{8}. For a spheroid, the asymptotic behavior of the impedance at high frequency is given by a term linear in frequency, with coefficient given by the sum of spheroid mass and added fluid mass, and a term proportional to the square root of frequency, with a Basset coefficient. Lawrence and Weinbaum \cite{4} calculated the latter from an expression derived by Landau and Lifshitz \cite{9} and Batchelor \cite{10} for a body of arbitrary shape. The Basset coefficient differs from the coefficient of the term linear in the square root of frequency at low frequency. In previous work we have calculated the Basset coefficient for a general ellipsoid for the three principal directions \cite{11}.

The steady state friction coefficient and the effective mass are sufficient for the construction of a simple, approximate expression for the complex admittance, showing two poles in the complex plane of the square root of frequency. On the basis of their numerical work Lawrence and Weinbaum suggested a three-pole expression for the complex admittance corresponding to the axial motion of a spheroid, involving the steady-state friction coefficient, the effective mass, and the Basset coefficient \cite{4}. It has been suggested that the approximate expression provides an adequate approximation to the actual admittance \cite{12}$^,$\cite{13}. In the following we examine this idea by comparison with a calculation of the admittance on the basis of the method of Pad\'e approximants \cite{14}. Our conclusion is that the simple two-pole expression is quite useful, and that the small deviations from the two-pole approximation are not well described by the Lawrence-Weinbaum expression.

\section{\label{II}Admittance}

We consider a prolate or oblate spheroid of mass $m_p$ immersed in a viscous incompressible fluid of shear viscosity $\eta$, mass density $\rho$. The fluid is assumed to fill the whole outer space, and a no-slip boundary condition holds at the surface of the spheroid. The spheroid and the fluid are at rest for $t<0$. We choose a cylindrical frame of coordinates $(r,\varphi,z)$ such that the center of the spheroid at rest is at the origin, and the $z$ axis is the axis of symmetry. Furthermore $r$ is the radial distance from the axis, and $\varphi$ is the azimuthal angle. We follow the convention of Happel amd Brenner \cite{15}, and denote the longest semi-axis by $a$ and the shortest by $b$ in both the prolate and the oblate case. At time $t=0$ a sudden impulse $\vc{S}=S\vc{e}_z$ in the direction of the symmetry axis is imparted to the spheroid, corresponding to the applied force
\begin{equation}
\label{1}\vc{F}(t)=S\vc{e}_z\delta(t).
\end{equation}
The impulse is assumed to be sufficiently small that the resulting flow velocity $\vc{v}(\vc{r},t)$ and pressure $p(\vc{r},t)$ can be assumed to satisfy the linearized Navier-Stokes equations
\begin{equation}
\label{2}\rho\frac{\partial\vc{v}}{\partial t}=\eta\nabla^2\vc{v}-\nabla p,\qquad\nabla\cdot\vc{v}=0.
\end{equation}
The flow velocity $\vc{v}$ tends to zero at infinity, and the pressure $p$ tends to a constant $p_{st}$. We are interested in the corresponding velocity $\vc{U}(t)=U(t)\vc{e}_z$ of the spheroid.

The equation of motion for the spheroid reads
\begin{equation}
\label{3}m_p\frac{dU}{dt}=K(t)+F(t),
\end{equation}
where $K(t)$ is the hydrodynamic force exerted by the fluid, which may be calculated from the fluid stress tensor. In Fourier transform the equation becomes
\begin{equation}
\label{4}-i\omega m_pU_\omega=K_\omega+F_\omega.
\end{equation}
Since $K_\omega$ is linear in $U_\omega$ this may be resolved to the form
\begin{equation}
\label{5}U_\omega=\mathcal{Y}(\omega)F_\omega,
\end{equation}
with admittance $\mathcal{Y}(\omega)$. The latter may be cast in the form
\begin{equation}
\label{6}\mathcal{Y}(\omega)=\big[-i\omega(m_p+m_a)+\zeta(\omega)\big]^{-1}.
\end{equation}
Here $m_a$ is the added mass, which follows from the high-frequency behavior. The friction coefficient $\zeta(\omega)$ is at most linear in $\sqrt{\omega}$ at high frequency. The impedance is the inverse of the admittance.

The added mass $m_a$ follows from the theory of potential flow, and is well known \cite{2}.
The friction coefficient at zero frequency follows from the solution of the steady state Stokes equations. It takes the form
\begin{equation}
\label{7}\zeta(0)=6\pi\eta R,
\end{equation}
where $R$ has the dimension length. Its explicit expression is also well known \cite{2}.

For a sphere of radius $a$ the friction coefficient takes the simple form \cite{7}
\begin{equation}
\label{8}\zeta(\omega)=6\pi\eta a(1+\alpha a),
\end{equation}
where $\alpha=\sqrt{-i\omega\rho/\eta},\;\mathrm{Re}\;\alpha>0$.
Similarly, for a spheroid the friction coefficient can be expanded in powers of $\alpha$ as
\begin{equation}
\label{9}\zeta(\omega)=\zeta^{(0)}+\zeta^{(1)}\alpha+\zeta^{(2)}\alpha^2+O(\alpha^3).
\end{equation}
The term linear in $\alpha$ can be expressed in terms of the coefficient at zero frequency $\zeta(0)=\zeta^{(0)}$ as
\begin{equation}
\label{10}\zeta^{(1)}=\frac{1}{6\pi\eta}\zeta(0)^2.
\end{equation}
A similar theorem in tensor form holds for a body of arbitrary shape \cite{5}$^,$\cite{6}.

The first term in the asymptotic expansion of $\zeta(\omega)$ in powers of $\alpha^{-1}$ is linear in $\alpha$,
\begin{equation}
\label{11}\zeta(\omega)=6\pi\eta\alpha a^2B+O(1),\qquad \mathrm{as}\;\alpha\rightarrow\infty,
\end{equation}
where $B$ is the dimensionless Basset coefficient \cite{16}$^,$\cite{4}. For a sphere $B=1$ from Eq. (8). The explicit expression for an ellipsoid can be evaluated from the theory of potential flow \cite{2} by use of an argument due to Landau and Lifshitz \cite{9} and to Batchelor \cite{10}. The expression for an oblate spheroid was derived by Lawrence and Weinbaum \cite{4}. It agrees with the one derived for a general ellipsoid \cite{11}. The corresponding expression for a prolate spheroid has also been derived \cite{11}$^,$\cite{12}.

It is convenient to write the admittance in the form
 \begin{equation}
\label{12}\mathcal{Y}(\omega)=\big[-i\omega m_p+\mathcal{Y}_0(\omega)^{-1}\big]^{-1},\qquad\mathcal{Y}_0(\omega)=\big[-i\omega m_a+\zeta(\omega)\big]^{-1},
\end{equation}
where $\mathcal{Y}_0(\omega)$ is related to the Fourier transform of the hydrodynamic force by
 \begin{equation}
\label{13}U_\omega=-\mathcal{Y}_0(\omega)K_\omega.
\end{equation}
It clearly is independent of the mass $m_p$.

We define the viscous relaxation time $\tau_v=a^2\rho/\eta$, and the corresponding complex variable $\lambda=\sqrt{-i\omega\tau_v}=\alpha a$. The function $\mathcal{Y}_0(\omega)$ may be cast in the form \cite{11}
\begin{equation}
\label{14}\mathcal{Y}_0(\omega)=\frac{1}{6\pi\eta}\frac{1}{R+(R^2/a)\lambda+M_0\lambda^2+a\lambda^2\psi(\lambda)},
\end{equation}
where $M_0=m_a/(6\pi a^2\rho)$, and the function $\psi(\lambda)$ tends to zero at large $\lambda$. For a sphere $R=a,\;M_0=a/9$, and $\psi(\lambda)=0$. In that case the expression for $\mathcal{Y}_0(\omega)$ has just two poles in the complex $\lambda$ plane.

More generally, the function $\psi(\lambda)$ may be written as a continued fraction
 \begin{equation}
\label{15}\psi(\lambda)=\frac{A_0}{1+\frac{A_1\lambda}{1+A_2\lambda...}}.
\end{equation}
A Pad\'e approximant to the exact function is obtained by breaking off at some level. A three-pole approximation to the admittance is found by putting
 \begin{equation}
\label{16}\psi_3(\lambda)=\frac{A_0}{1+A_1\lambda}
\end{equation}
with suitably chosen coefficients $A_0,\;A_1$. In order to agree with the high-frequency behavior given by Eq. (11) we must have
 \begin{equation}
\label{17}\frac{A_0}{A_1}=B-\frac{R^2}{a^2}
\end{equation}
in the three-pole approximation. Lawrence and Weinbaum \cite{4} have suggested on the basis of their explicit calculations for a spheroid that a good approximation to the admittance $\mathcal{Y}_0(\omega)$ is found by use of Eq. (16) with $A_1=1$ and the coefficient $A_0$ calculated from the Basset coefficient $B$ by use of Eq. (17). In the following we examine this suggestion for a test case, for both an oblate and a prolate spheroid.

\section{\label{III}Hydrodynamic force}

The calculation will be based on the solution of the linearized Navier-Stokes equations Eq. (2) after a Fourier transform in time. By axial symmetry the flow velocity and the pressure can be derived from a scalar Stokes stream function. The partial differential equation for the stream function separates in spheroidal coordinates, and the solution can be expressed as an infinite sum of products of radial and angular spheroidal wavefunctions \cite{4}. In the calculation we have used spheroidal wavefunctions as programmed in Mathematica 6.

It has been shown by Lawrence and Weinbaum \cite{16} that the hydrodynamic force $K_\omega$ can be evaluated from the behavior of the stream function at large distance from the origin, i.e. from the value of the lowest order multipole moment.  The exact set of coupled multipole equations has been derived by Lai and Mockros \cite{3}. It was shown by Lawrence and Weinbaum \cite{4} that the infinite set of linear coupled multipole equations can be solved approximately by truncation and matrix inversion.

It is convenient to define the dimensionless reduced hydrodynamic force $G(\omega)$ by
\begin{equation}
\label{18}G(\omega)=K_\omega/K_0,\qquad G(0)=1.
\end{equation}
It follows from Eq. (13) that
\begin{equation}
\label{19}G(\omega)=6\pi\eta R\mathcal{Y}_0(\omega).
\end{equation}
We denote the corresponding expression obtained by putting $\psi(\lambda)=0$ in Eq. (14) as
\begin{equation}
\label{20}G_{2p0}(\omega)=\frac{R}{R+(R^2/a)\lambda+M_0\lambda^2}.
\end{equation}
This is a two-pole approximation to the actual function. We call this a two-pole approximation, since the function has two poles in the complex $\sqrt{\omega}$ plane. The deviation from the exact function is characterized by the function $\psi(\lambda)$ in Eq. (14).

We have used the exact multipole equations to calculate the function $\psi(\lambda)$ at a chosen set of positive values $\lambda_j$ corresponding to values on the positive imaginary axis of the complex frequency plane. As test cases we have chosen a prolate and an oblate spheroid with aspect ratio $b/a=0.5$. It follows from Table 4-26.1 in Happel and Brenner \cite{15} that the corresponding value of the steady-state friction coefficient is $\zeta(0)=0.9053\times 6\pi\eta a$ in the oblate case, and $\zeta(0)=1.2039\times 6\pi\eta b$ in the prolate case. For values of the aspect ratio $b/a$ closer to unity the correction $\psi(\lambda)$ is very small, and hardly worth calculating. For values of the aspect ratio much less than unity a large number of multipoles must be included to achieve convergence, and the numerical calculation becomes more demanding.

In Table I we list the values $\psi_j=\psi(\lambda_j)$ at the chosen values $\lambda_1=0.2,\;\lambda_2=0.4,\;\lambda_3=0.7,\;\lambda_4=1,\;\lambda_5=1.5,\;\lambda_6=2$, and compare with the corresponding values of the function
\begin{equation}
\label{21}\psi_{LW}(\lambda)=\frac{B-R^2/a^2}{1+\lambda},
\end{equation}
corresponding to the approximation of Lawrence and Weinbaum \cite{4}. For the oblate spheroid there is fair agreement, but for the prolate spheroid the LW-approximation yields values with the wrong sign. The numerator in Eq. (21) is negative for the prolate spheroid. It is positive for an oblate spheroid.

\section{\label{IV}Velocity relaxation function}

In this section we derive approximate expressions for the velocity relaxation function based on the results derived above, and show numerical results for a prolate and an oblate spheroid of aspect ratio $b/a=0.5$. By inversion of the Fourier transform in Eq. (5) we find
 \begin{equation}
\label{22}U(t)=\Psi(t)S,\qquad t>0.
\end{equation}
The initial value of the velocity relaxation function $\Psi(t)$ is
 \begin{equation}
\label{23}\Psi(0+)=\frac{1}{m^*},\qquad m^*=m_p+m_{a},
\end{equation}
with effective mass $m^*$. The initial value is less than $1/m_p$ due to loss of momentum to infinity via longitudinal sound waves \cite{17}. The low frequency behavior of the friction tensor, given by Eq. (9), corresponds to the long-time behavior \cite{18}
 \begin{equation}
\label{24}\Psi(t)\approx\frac{\sqrt{\rho}}{12(\pi\eta t)^{3/2}},\qquad \mathrm{as}\;t\rightarrow\infty,
\end{equation}
independent of the shape and size of the spheroid and its mass density $\rho_p$.

We write the relaxation function in the form
 \begin{equation}
\label{25}\Psi(t)=\frac{1}{m^*}\gamma(t),
\end{equation}
with dimensionless function $\gamma(t)$ with initial value $\gamma(0)=1$. The function has the one-sided Fourier transform
 \begin{equation}
\label{26}\Gamma(\omega)=\int^\infty_0e^{i\omega t}\gamma(t)\;dt=m^*\;\mathcal{Y}(\omega).
\end{equation}
The simplest estimate of the function $\gamma(t)$ is the exponential $\exp(-t/\tau_M)$, with mean relaxation time $\tau_M=m^*/(6\pi\eta R)$, but this violates the long-time behavior given by Eq. (24).

An approximate relaxation function which accounts for the initial value, the mean relaxation time $\tau_M$, and the long-time behavior (24), is given by the approximate two-pole expression
 \begin{equation}
\label{27}\Gamma_{2p}(\omega)=\frac{a^2\rho}{\eta}\;\frac{M}{R+(R^2/a)\lambda+M\lambda^2},
\end{equation}
where $M=m^*/(6\pi a^2\rho)$ has the dimension length.  In the two-pole approximation the function $\gamma(t)$ takes the approximate form \cite{18}
 \begin{equation}
\label{28}\gamma_{2p}(t)=\frac{1}{\sqrt{\sigma^2-4}}\;\big[v_+w(-iv_+\sqrt{t/\tau_M})-v_-w(-iv_-\sqrt{t/\tau_M})\big],
\end{equation}
where
 \begin{equation}
\label{29}\sigma=\sqrt{\frac{6\pi\rho R^3}{m^*}},\qquad v_\pm=-\frac{1}{2}\sigma\pm\frac{1}{2}\sqrt{\sigma^2-4},
\end{equation}
and $w(z)=\exp(-z^2)\mathrm{erfc}(-iz)$.

The two-pole approximation is determined by the effective mass and the friction coefficient at zero frequency. In order to find the actual behavior of the velocity relaxation function, we must take account of the function $\psi(\lambda)$, defined in Eq. (14). The approximation of Lawrence and Weinbaum, corresponding to Eq. (21), leads to a three-pole approximation to the velocity relaxation function, which we denote as $\gamma_{LW}(t)$. More generally, the time-dependence of the reduced velocity relaxation function in an $N$-pole approximation is given by \cite{14}
 \begin{equation}
\label{30}\gamma_N(t)=\sum^N_{k=1}R_kv_kw(-iv_k\sqrt{t/\tau_v}),
\end{equation}
where $\{v_k\}$ are the $N$ poles of the one-sided Fourier transform $\Gamma_N(\omega)$ in the complex $\lambda$ plane, and $\{R_k\}$ are the residues.

The function $\Gamma_N(\omega)$ is found numerically from a Pad\'e approximant to the function $\psi(\lambda)$ determined by the values of the reduced hydrodynamic force $G(\omega)$ at selected positive values $\{\lambda_j\}$. For example, the six values listed in Table I are sufficient to determine a 5-pole approximation $\gamma_5(t)$. In this case the Pad\'e approximant to the function $\psi(\lambda)$ is the ratio of a cubic and a quintic polynomial in $\lambda$.

We find numerically that for aspect ratio $b/a=0.5$ it suffices to include multipoles of order $n=17$, in the notation of Lawrence and Weinbaum \cite{4}. This corresponds to a set of nine coupled multipole equations, since only odd multipoles occur. We have calculated the values of $\psi(\lambda)$ at eight positive values $\{\lambda_j\}$, corresponding to a six-pole approximation $\gamma_6(t)$. It turns out that in fact two of the residues are negligibly small, so that $\gamma(t)$ is represented accurately by four pole contributions. The larger number of values $\{\lambda_j\}$ allow a more accurate determination of the relevant residues and poles. The difference between the approximations $\gamma_5(t)$ and $\gamma_6(t)$ cannot be seen graphically.

In Fig. 1 we plot the ratio $\gamma_6(t)/\gamma_{2p}(t)$ as a function of $\log_{10}(t/\tau_v)$ for an oblate spheroid with aspect ratio $b/a=0.5$ and vanishing mass density $\rho_p=0$, and compare with the approximation of Lawrence and Weinbaum calculated from Eq. (21). In Fig. 2 we present the corresponding plots for a neutrally buoyant spheroid with $\rho_p=\rho$. In Figs. 3 and 4 we present similar plots for a prolate spheroid with aspect ratio $b/a=0.5$ and mass densities $\rho_p=0$ and $\rho_p=\rho$, respectively. It turns out that for the oblate spheroid the Lawrence-Weinbaum approximation agrees qualitatively, but that for the prolate spheroid it is qualitatively wrong. In both cases the deviation from the two-pole approximation $\gamma_{2p}(t)$ is less than one percent.

\section{\label{VI}Discussion}

The frequency-dependent admittance of a prolate or oblate spheroid can be analyzed efficiently by the method of Pad\'e approximants applied to the approximate solution of the exact set of coupled equations for the multipole moments of the hydrodynamic force density. The method leads to an approximate representation of the admittance as a ratio of two polynomials in the square root of frequency. Hence one can evaluate the time-dependent velocity relaxation function as a sum of $w$ functions.

We have applied the method to a prolate or oblate spheroid of aspect ratio $b/a=0.5$ for two values of the spheroid mass density, and have compared the numerical results with those calculated from the approximation proposed by Lawrence and Weinbaum \cite{4}. The comparison is shown in Figs. 1-4. The figures show that for the cases considered the deviation from the two-pole approximation is quite small, and not well described by the approximation of Lawrence and Weinbaum \cite{4}, based on just the Basset coefficient.

The elaborate calculation in terms of spheroidal wavefunctions leads to only a small deviation from the result which may be calculated from the steady state friction coefficient and the effective mass, as given by the classical expressions for a general ellipsoid \cite{2}$^,$\cite{11}. Presumably an exact calculation for an ellipsoid would lead to the same conclusion in that case. For many practical purposes \cite{19} the two-pole approximation to the admittance, as determined from steady state friction coefficient and effective mass, will be sufficient. The use of only the Basset coefficient for a calculation of the deviations from the two-pole approximation on the intermediate timescale is questionable.

\newpage
\begin{table}[!htb]  \footnotesize\centering
  \caption{}\label{tab:1}
\begin{tabular}{|c|c|c|c|c|c|c|c|}
\hline
 $$ &$\lambda$ &$0.2$& $0.4$ &$0.7$ &$1$ &$1.5$ &$2$
 \rule[-5pt]{0pt}{16pt} \\
\hline
 $\mathrm{oblate}$ &$10^5\psi$ &$521$ & $473$ &$409$ &$358$ &
$294$&$247$
 \!\!\!\rule[-5pt]{0pt}{16pt} \\
 $$ &$10^5\psi_{LW}$ &$511$ & $438$ &$361$ &$307$ &
$245$&$204$
 \!\!\!\rule[-5pt]{0pt}{16pt} \\
  $\mathrm{prolate}$ &$10^5\psi$ &$58$ & $45$ &$27$ &$15$ &
$ 1$&$-8$
 \!\!\!\rule[-5pt]{0pt}{16pt} \\
  $$ &$10^5\psi_{LW}$ &$-248$ & $-212$ &$-175$ &$-149$ &
$-119$&$-99$
 \!\!\!\rule[-5pt]{0pt}{16pt} \\
\hline
\end{tabular}
\end{table}

\section*{Table caption}
Table of values of the function $\psi(\lambda)$ at selected values of $\lambda$ for an oblate and a prolate spheroid of aspect ratio $b/a=0.5$, as calculated from the set of coupled multipole equations, compared with the Lawrence-Weinbaum approximation given by Eq. (21).
\newpage

\subsection*{Fig. 1}
Plot of the ratio $\gamma_6(t)/\gamma_{2p}(t)$ as a function of $\log_{10}(t/\tau_v)$ for an oblate spheroid with aspect ratio $b/a=0.5$ and mass density $\rho_p=0$ (solid curve), compared with the corresponding ratio $\gamma_{LW}(t)/\gamma_{2p}(t)$, as given by the Lawrence-Weinbaum approximation Eq. (21) (dashed curve).
\subsection*{Fig. 2}
Plot of the ratio $\gamma_6(t)/\gamma_{2p}(t)$ as a function of $\log_{10}(t/\tau_v)$ for an oblate spheroid with aspect ratio $b/a=0.5$ and mass density $\rho_p=\rho$ (solid curve), compared with the corresponding ratio $\gamma_{LW}(t)/\gamma_{2p}(t)$, as given by the Lawrence-Weinbaum approximation Eq. (21) (dashed curve).
\subsection*{Fig. 3}
Plot of the ratio $\gamma_6(t)/\gamma_{2p}(t)$ as a function of $\log_{10}(t/\tau_v)$ for a prolate spheroid with aspect ratio $b/a=0.5$ and mass density $\rho_p=0$ (solid curve), compared with the corresponding ratio $\gamma_{LW}(t)/\gamma_{2p}(t)$, as given by the Lawrence-Weinbaum approximation Eq. (21) (dashed curve).
\subsection*{Fig. 4}
Plot of the ratio $\gamma_6(t)/\gamma_{2p}(t)$ as a function of $\log_{10}(t/\tau_v)$ for a prolate spheroid with aspect ratio $b/a=0.5$ and mass density $\rho_p=\rho$ (solid curve), compared with the corresponding ratio $\gamma_{LW}(t)/\gamma_{2p}(t)$, as given by the Lawrence-Weinbaum approximation Eq. (21) (dashed curve).

\newpage
\setlength{\unitlength}{1cm}
\begin{figure}
 \includegraphics{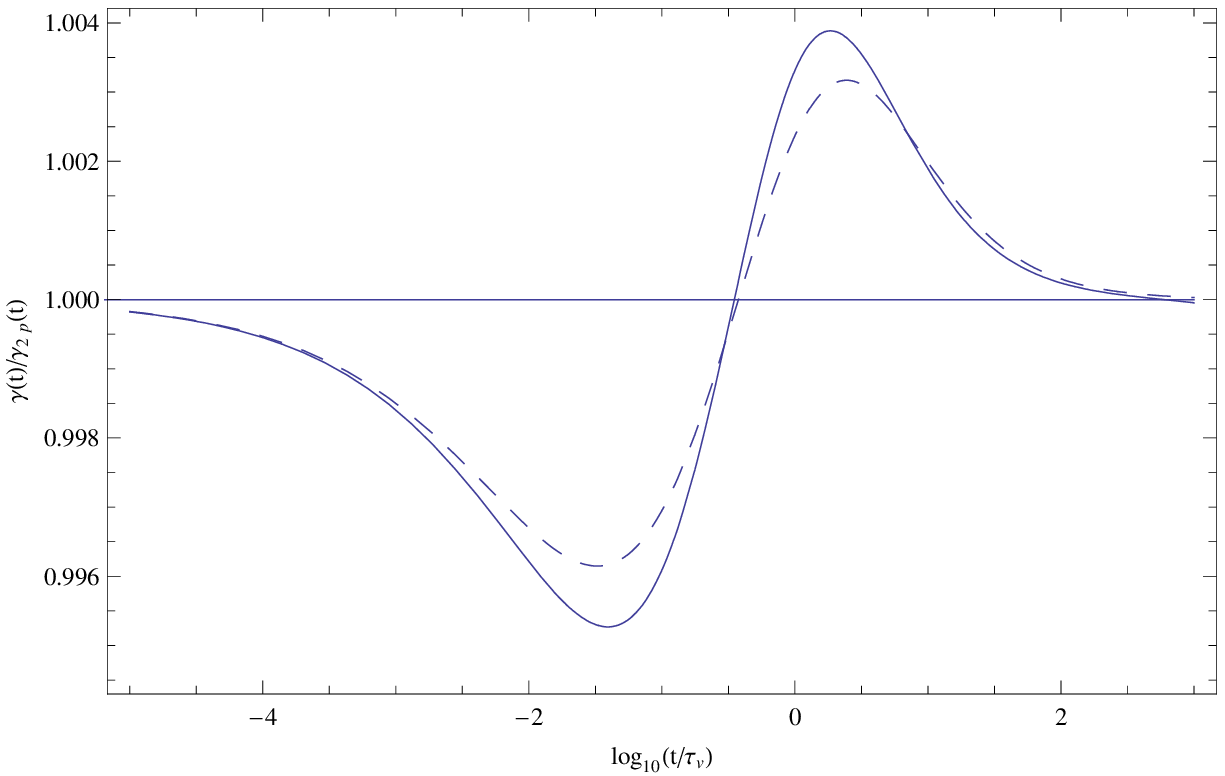}

  \caption{}
\end{figure}
\newpage
\clearpage
\newpage
\setlength{\unitlength}{1cm}
\begin{figure}
 \includegraphics{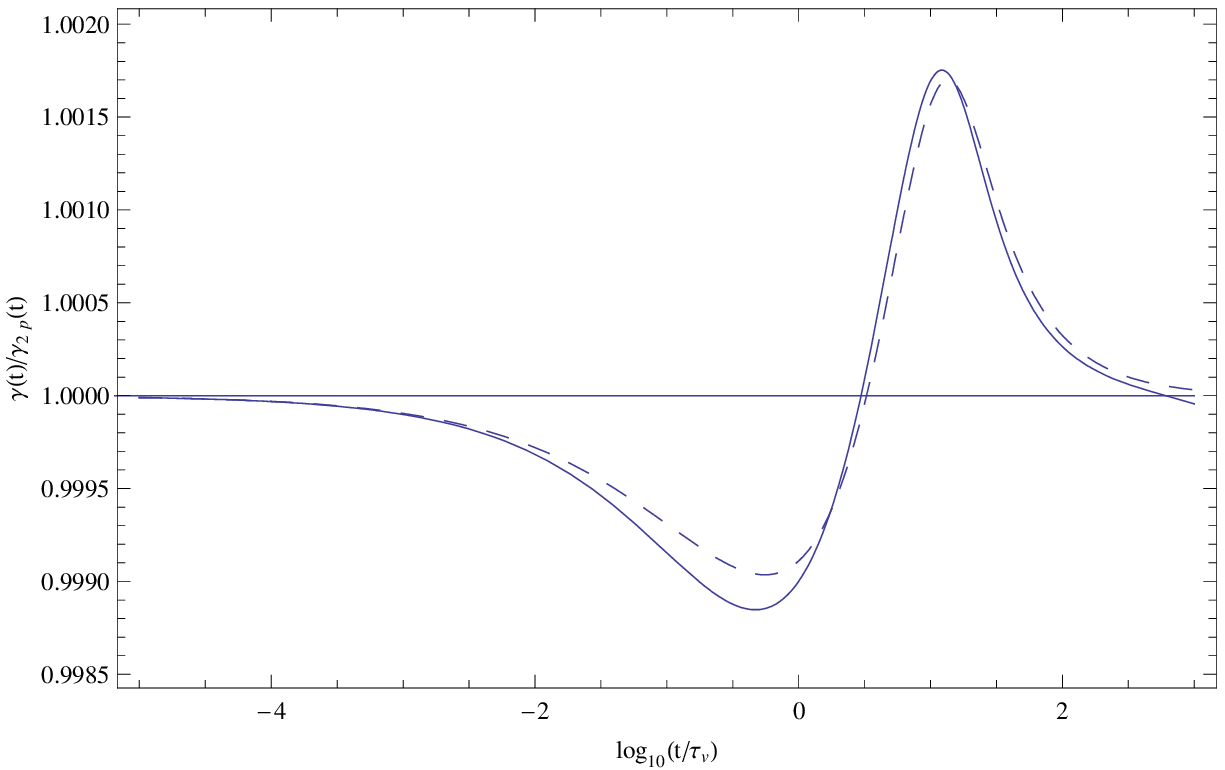}

  \caption{}
\end{figure}
\newpage
\clearpage
\newpage
\setlength{\unitlength}{1cm}
\begin{figure}
 \includegraphics{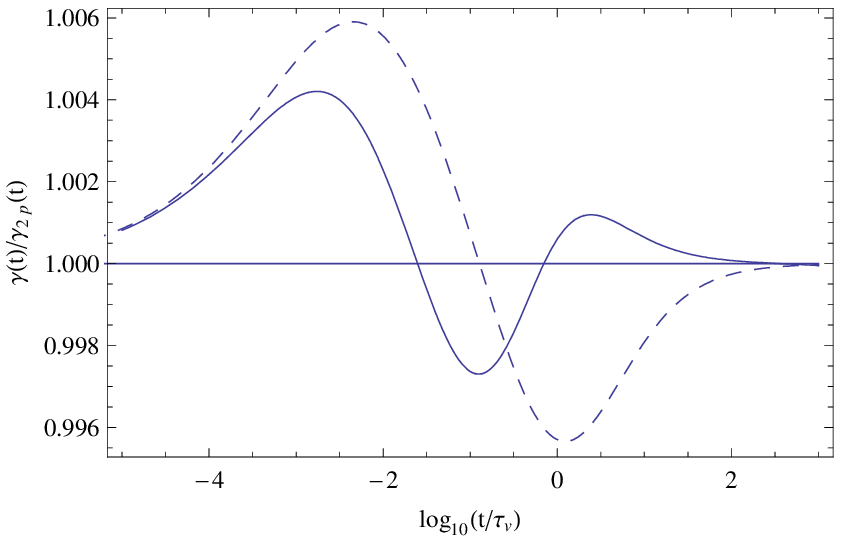}

  \caption{}
\end{figure}
\newpage
\clearpage
\newpage
\setlength{\unitlength}{1cm}
\begin{figure}
 \includegraphics{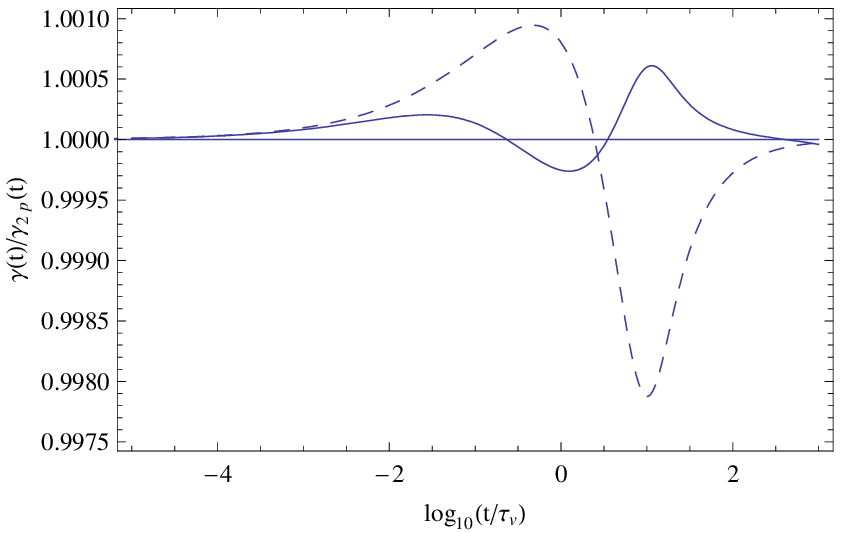}

  \caption{}
\end{figure}

\end{document}